\pdfoutput=1
\documentclass[%
  prb,%
  letterpaper,%
  twocolumn,%
  showpacs,%
  floatfix,%
  superscriptaddress%
]{revtex4}

\RequirePackage{graphicx,amsmath,amssymb,bm}
\RequirePackage{hyperref}
\hypersetup{%
  breaklinks = {true},
  citecolor = {blue},
  colorlinks = {true},
  linkcolor = {red},
  pdfauthor = {\textcopyright\ Vitor M. Pereira},
  pdfcreator = {\LaTeX\ and \flqq hyperref\frqq},
  pdffitwindow = {true},
  pdfmenubar = {true},
  pdfpagelayout = {SinglePage},
  pdfstartview = {Fit},
  pdftoolbar = {true},
  plainpages = {false},
}

\newcommand{\Fref}[1]{Fig.~\ref{#1}}
\newcommand{\Eqref}[1]{Eq.~(\ref{#1})}
\renewcommand{\eqref}[1]{eq.~(\ref{#1})}

\RequirePackage[usenames,dvipsnames]{xcolor}
\usepackage{soul}

\begin{document}

\title{Conductance across strain junctions in graphene nanoribbons}

\author{D. A. Bahamon}
\affiliation{Graphene Research Centre and
            Department of Physics, National University of Singapore,
            2 Science Drive 3, Singapore 117542}

\author{Vitor M. Pereira}
\email[Corresponding author: ]{vpereira@nus.edu.sg}
\affiliation{Graphene Research Centre and
            Department of Physics, National University of Singapore,
            2 Science Drive 3, Singapore 117542}

\date{\today}

\begin{abstract}
To address the robustness of the transport gap induced by locally strained 
regions in graphene nanostructures, the effect of disorder and smoothness of the 
interface region is investigated within the Landauer-B\"uttiker formalism. The 
electronic conductance across strained junctions and barriers in graphene 
nanoribbons is calculated numerically, with and without various types of 
disorder, and comparing smooth and sharp strain junctions. A smooth strain 
barrier in graphene is seen to be generically as efficient in suppressing 
transport at low densities as a sharp one, and the critical density (or 
energy) for the onset of transmission depends on the strain orientation
with respect to the ribbon. In addition, hopping (or strain) inhomogeneity and 
work function mismatch at the interface region do not visibly degrade the 
transport gap. These results show that the strain-induced transport gap at a 
strain junction is robust to more realistic strain conditions.
\end{abstract}




\maketitle


\section{Introduction}

The ability to control the flow of charge through nanoscale devices --- in 
particular being able to efficiently establish clearly defined on and off 
states --- is one of the perennial driving forces and goals of research towards 
electronic devices of smaller dimensions or with new functionalities. 
The advent of graphene has brought an entire new realm of possibilities in this 
front \cite{Schwierz_GFET}. The intrinsic two-dimensionality of graphene and a 
large number of other emerging strictly 2D crystals means that they can be 
readily integrated with the 
conventional 2D device fabrication paradigm in electronics 
\cite{Novoselov_CN2D}. But, more importantly, being a pure surface opens several 
new possibilities of 
significantly modifying their electronic structure and transport properties, 
such as by tailoring the interaction with substrates, with the chemical 
environment, with electromagnetic radiation, or doping. In the case of 
graphene, in particular, having the strongest covalent bonding in nature leads 
to its record high tensional modulus of $\sim$1\,TPa, while, at the same time, 
being able to withstand planar and elastic deformations as high as 15--20\,\% 
\cite{Cadelano:2009,leeSCIENCE2008}. Hence, unlike more conventional condensed 
matter systems whose intrinsic 
brittleness rarely allows elastic deformations of even the order of 1\,\% 
\cite{flex_electronics}, large mechanical deformations in graphene are a reality 
\cite{LevySci2010,LuNC2012,Tomori} and a potential new means for external 
control of its electronic properties \cite{VPereira3,VPereira2,GuineaNP2010}. 
Its extraordinary mechanical robustness further allows the exploration of the 
third dimension, with the potential for manipulating this electronic membrane 
with strain and curvatures tailored for particular functionalities 
\cite{FoglerScroll,KimFolds,RainisFolds}.

Apart from these practical and functional advantages of having a strong 
crystalline metallic membrane, the coupling of lattice deformations to electrons 
in graphene is also of great interest from a fundamental point of view. The 
nature of the honeycomb lattice has the consequence that, in addition to the 
conventional coupling through the so-called deformation potential, electrons in 
graphene feel the local lattice deformations through an additional 
pseudomagnetic vector potential coupling \cite{Kane:1997,Suzuura:2002}. The 
direct consequence is that local non-uniform deformations translate directly 
into an 
effective pseudomagnetic field \cite{netoRMP2009,yo4}. This fact, realized 
theoretically long ago, has been recently established with the experimental 
observation of the reconstruction of the Dirac spectrum into Landau-levels in 
regions of locally strained graphene \cite{LevySci2010,LuNC2012}. The magnitude 
of 
the effective pseudomagnetic fields extracted from these experiments can easily 
fall into the 300--600\,Tesla range, vigorously attesting to the strong impact 
that mechanical deformations can have in the electronic structure of graphene.

Against the backdrop of these experiments and these numbers, the prospect of 
strain-engineering graphene's intrinsic electronic response gains more traction, 
as reflected by the increased interest in exploring and characterizing strained 
graphene electrically, optically, or magnetically 
\cite{Pereira-OpticalStrain:2010,Pellegrino:2010,Bunch26012007,Piezoelectricity, 
LowStrain}. One of the simplest ideas whereby local strain can be used to modify 
the transport characteristics of a graphene channel consists of a linear 
interface separating two regions of graphene in different states of uniaxial 
strain, thus establishing a one-dimensional strain junction or step 
\cite{VPereira3,FoglerJunction,PeetersStrain,PeetersStrain2}. Similar ideas 
have been proposed for other semiconducting systems such as Si \cite{strainSi} 
and, of course, films of heteroepitaxial Si have been instrumental in transistor 
mobility optimization \cite{StrainedSi}. 

In graphene such a simple interface and its variations in the form of barriers 
or superlattices has the potential to generate a transport gap at low densities 
\cite{FoglerJunction} (which is to say that the dependence of the electronic 
conductivity, $\sigma$, on the electron density, $n_e$, is modified such that 
$\sigma(n_e<n^*) = 0$), as well as to spatially confine some electronic modes to 
the barrier region \cite{VPereira3,Bao2009562,Kronig-Penney_strain}. If strain 
is uniform, albeit different, on both sides of the interface, a basic 
qualitative picture arises that allows us to understand the underlying physics. 
It can be summarized in that the associated pseudomagnetic potential vector will 
be different in the two regions defined by the interface, with the consequence 
that the Fermi circles for electrons on those two regions will appear centered 
at different points in the absolute (undeformed) reciprocal space. If momentum 
conservation along the interface direction is assumed, this displacement of the 
Fermi circle reduces the phase space available for transmission across the 
interface, which is entirely suppressed when the density is so low that the 
displacement is larger than the Fermi momentum \cite{VPereira3}.

This paper aims at analyzing and quantifying the robustness of the main 
transport feature --- the transport gap --- when the idealized conditions under 
which these strain barriers have been previously studied are relaxed. In 
particular, we will be interested in the fate of the transport gap when the 
strain interface becomes smooth, rather than the idealized step-like strain 
step. In addition to this necessary generalization, we shall also look at the 
effect of disorder at the interface. Electronic disorder can arise as a result 
of inhomogeneous and random strain at, and within some distance from, the 
interface, or as a result of impurities or enhanced interactions with the 
substrate at the interface region, depending on the strategy used to establish 
the strain step. In order to approach these situations in both an unbiased and 
unified way, the conductance of graphene through such strain steps 
and barriers is calculated directly on the lattice, numerically, within the 
Landauer-B\"uttiker formalism, where the energy-dependent transmission 
probability is calculated with the lattice Green's function technique.

Our calculations based on a graphene nanoribbon ribbon (GNR) geometry show that 
the strain-induced transport gap is robust under these more realistic strain 
barrier situations. More specifically, barrier smoothness and hopping (or 
strain) disorder have no impact in the presence or magnitude of the transport 
gap. In fact, a ``smooth'' strain barrier is 
``better'' than a ``sharp'' one in that it considerably improves the quality of 
conductance quantization. This is quite the opposite of what happens for an 
electrostatic step or barrier in graphene, where a smooth interface region leads 
to qualitative changes in the transmission properties, such as the 
considerable suppression of the (Klein) tunneling amplitude except for 
precisely normal incidence \cite{Falko}. Local potential disorder is seen 
to have only a minor effect if it is short-ranged, and long-range is the only 
model of disorder explored here that has a noticeable detrimental effect. 

Despite our utilization of the conductance extracted for finite-width 
GNRs, some of our conclusions should transfer directly and unchanged to
the thermodynamic limit, notably the robustness with respect to the barrier 
smoothness. At any rate, the effort towards bottom-up synthesis of GNRs 
with atomically precise edges has already lead to various successful strategies 
\cite{CNT2GNR,GNR_mol_precursors,WeeBottomUp}, and hence the prospect of narrow 
GNRs free 
from edge roughness is quite real, therefore allowing the direct testing of our 
calculations in that case. For example, the dependence of the Fano factor in 
strain magnitude could be used experimentally to quantify the strength of 
deformation in the sample. 

We emphasize from the outset that, here, we are not interested in the modulation 
of the intrinsic finite-size gap and the transition between metallic and 
insulating character of perfect GNRs that has been amply discussed by other 
authors \cite{ZGR_strain,Li,ZGR_strain3,Sun}. 

Even though direct signatures of strain in the electronic structure of graphene 
have been detected by Raman spectroscopy 
\cite{Raman_strainAG,Raman_strainNF,Raman_strainZN} and local probe 
spectroscopy \cite{LevySci2010,LuNC2012,YanPseudoLL,BockrathPseudoLL}, its 
effect on the electronic 
transport has not been experimentally investigated yet. We trust that the 
current drive to utilize graphene in flexible electronics, where mechanical 
deformation is inevitable to some extent, as well as our current results 
establishing the robustness of the transport gap, will motivate direct studies 
of the transport characteristics under strain.

\begin{figure}[t]
  \centering
  \includegraphics[width=8.5cm]{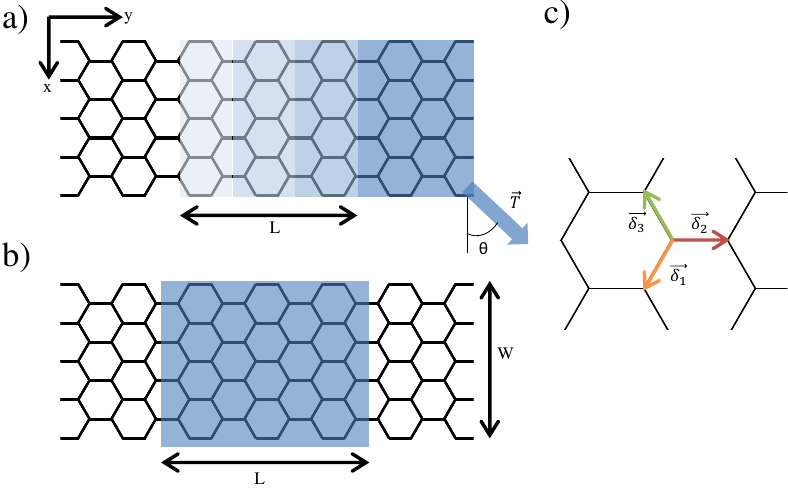}
   \caption{
    (Color online) Schematic representation of the 
    unstrained-strained junction (a) and of the strain barrier (b).
    (c) Honeycomb lattice with the convention used for the nearest-neighbor 
    vectors $\boldsymbol{\delta_1}=(\sqrt{3}a/2,- a/2)$, 
    $\boldsymbol{\delta_2}=(0,a)$, and
    $\boldsymbol{\delta_3}=(-\sqrt{3}a/2,-a/2)$.
  }
  \label{fig:fig1}
\end{figure}

\section{Model and Methodology}

The dynamics of electrons in graphene is modeled here as being 
described by a nearest-neighbor $\pi$-band tight-binding Hamiltonian for an 
hexagonal lattice,
\begin{equation}
H =   \sum_{<i,j>}t_{ij} (c_{i}^{\dagger} c_{j} +
c_{j}^{\dagger} c_{i})
,
\end{equation}
where $c_{i}$ represents the fermionic annihilation operator on site
$i$, $t_{ij}$ is the hopping amplitude between nearest neighbor $p_z$
orbitals (in the unstrained lattice $t_{ij}=t_0 \approx -2.7$\,eV). The
leading effect of uniaxial strain to the $pi$-band electrons arises through 
the modification of the carbon bond distances in equilibrium, $\bm{\delta}_i^0$ 
[cf. \Fref{fig:fig1}(c)], according to $\bm{\delta}_{i} = 
(1+\bm{\nabla u}) \cdot \bm{\delta^0_{i}} $, where $\bm{\nabla u}$ is the 
plane displacement gradient tensor. Since we are interested in 
characterizing the effects on the quantum transport 
across non-strained-to-strained junctions, throughout this work we shall 
consider only the representative case of uniaxial and planar deformations. In 
this case the deformation gradient tensor can be exactly replaced by its 
symmetric decomposition, the small strain tensor: $\bm{\varepsilon} = \bm{\nabla 
u}/2 + \bm{\nabla u}^\top/2$. In the 
coordinate system shown in in \Fref{fig:fig1} the strain tensor 
reads \cite{VPereira2}
\begin{equation}
  \boldsymbol{\varepsilon} = \varepsilon 
  \begin{pmatrix}
  \cos^2\theta-\sigma \sin^2\theta & (1+\sigma)\cos \theta \sin \theta  \\
  (1+\sigma)\cos \theta \sin \theta & \sin^2\theta-\sigma \cos^2\theta
  \end{pmatrix},
\end{equation}
where $\varepsilon$ is the magnitude of the applied tension, $\theta$ is the 
orientation with respect to the x direction (see \Fref{fig:fig1}), and 
$\sigma=0.165$ is the Poisson ratio.
The strain-induced changes in the three nearest neighbor vectors 
$\boldsymbol{\delta_i}$ modify the value of the corresponding hopping 
amplitudes. Their dependence on the distance between neighboring $p_z$ orbitals 
is modeled to vary according to $t_i=t_0e^{-3.37(\delta_i/a-1)}$ 
\cite{VPereira2}, with 
$a=\delta^0=0.142$\,nm being the unstrained C-C bond length. 

Of special significance are the cases of strain applied along the 
lattice zigzag (e.g., $\theta=0$) and armchair (e.g., $\theta=\pi/2$) directions 
\cite{VPereira2}. These two particular tension directions leave $t_1=t_3$ in 
both cases, and the nearest-neighbor distances are modified explicitly as
\begin{subequations}
\begin{align}
  \theta=0: \qquad &
|\delta_1|=|\delta_3|=\bigl(1+\tfrac{3\varepsilon}{4}-\tfrac{\varepsilon\sigma} 
 {4}\bigr)a \\
  & |\delta_2|=(1-\varepsilon\sigma)a \\
  \intertext{and}
  \theta=\tfrac{\pi}{2}: \qquad &
|\delta_1|=|\delta_3|=\bigl(1+\tfrac{1\varepsilon}{4}-\tfrac{3\varepsilon\sigma}
  { 4 } \bigr)a \\
  & |\delta_2|=(1+\varepsilon)a
\end{align}
\end{subequations} 

Effects of strain in the spectrum of GNRs have been considered elsewhere by 
various authors. But these studies pertain mostly to situations where the 
entire nanostructure is under strain, and not the cases of strain barriers or 
junctions considered here. For example, it has been shown that strain does 
not qualitatively affect the intrinsic bandstructure of zigzag GNRs
\cite{ZGR_strain,Li,ZGR_strain3}, whereas band structure calculations 
for AGNRs have shown a variable intrinsic gap with a sawtooth shape as 
a function of the applied strain. However, in the latter case the value of 
the intrinsic gap still scales with the inverse width of the ribbon, 
and hence becomes rapidly insignificant as the transverse dimension
increases \cite{Sun}. For calculational convenience, our analysis of the 
strain-induced transport gap generated by strain junctions will focus only on 
AGNRs, since these allow an analytical closed form for the propagating modes, 
which will be an important factor in the interpretation and analysis of the 
results. With reference to \Fref{fig:fig1}, ribbons of different width are 
identified by the number of atoms along one of the vertical zigzag chains, 
$\mathcal{W}$. Moreover, given that we are interested in following the 
evolution of the transport gap, the incoming (and outgoing) contact in the case 
of the junction (barrier) is always metallic. This means that 
we shall always choose $\mathcal{W}=3m+2$, where $m$ is a positive integer. The 
width of the nanoribbon is denoted by $W$, and is given by 
$W={\sqrt{3}a(\mathcal{W}-1)}/2$.

In the case of the unstrained-strained junction we define three regions in the 
infinite nanoribbon: an unstrained semi-infinite contact acting as a waveguide 
for the incident electrons, a strained semi-infinite contact acting as a 
waveguide for the transmitted electrons, and a central region of length $L$ 
acting as the scattering region (in this work we consider $L$ equal to the 
width $W$ of the GNR). The transition from the unstrained to the strained 
region is not abrupt. Instead, strain is smoothly incremented over a length 
$L_s$ ($L_s < L$), which is a parameter that we can vary to assess the 
influence of the barrier smoothness on the conductance suppression at low 
densities. This is achieved with the following position dependent strain 
\begin{equation}
  \varepsilon(y)=\frac{\varepsilon_0}{e^{-(y-L/2)/L_s} + 1},
\label{eq:smooth}
\end{equation}
where $y=0$ and $y=L$ mark the end of the unstrained left and the beginning of
the right strained contact, respectively. 
For the study of the strain barrier, the infinite ribbon is likewise divided in 
three regions: two semi-infinite unstrained contacts and a strained central 
region of length $L=W$. The strain magnitude is also smoothed at the interfaces 
of the central region with the contacts as described above.

The conductance of the graphene nanostructures is evaluated within the 
Landauer-B$\ddot{u}$ttiker formalism, whereby $G(E)=G_0\,T(E)$, with
$G_0=\frac{2e^2}{h}$. $T(E)$ is the transmission function, and is obtained 
using the recursive lattice Green's function technique \cite{Lewenkopf_resumo}.

\section{Conductance of an unstrained-strained junction}

\begin{figure}[t]
  \centering
  \includegraphics[width=8.0cm]{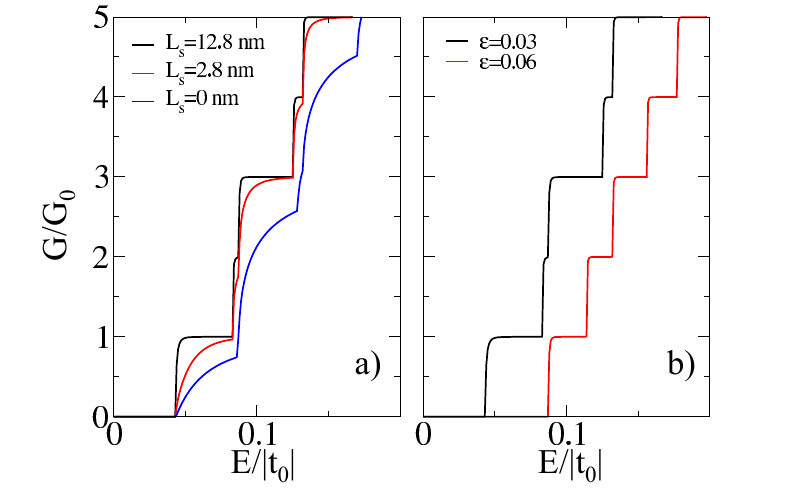}
\caption{
    (Color online) (a) Conductance of an unstrained-strained junction with
    $W=15.2$\,nm, $\varepsilon=0.03$, $\theta=\pi/2$, and different values of 
    the smoothing length, $L_s$. Notice that the plateaux become increasingly 
    well defined with the introduction of a smoothing length $L_s\ne 0$, 
    but the fundamental suppression of conductance at low energies remains 
    unaffected.
    (b) Conductance for the same system with $L_s=12.8$\,nm at different 
    strain magnitudes.
  }
  \label{fig:Gjunction}
\end{figure}

In order to understand how strain modifies the conductance of an AGNR and to 
compare it with the results obtained within the continuum approach based on the 
Dirac equation \cite{VPereira3,Fogler,Pellegrino} let us begin by describing 
the quantum transport features of the unstrained-strained junction. 
\Fref{fig:Gjunction}(a) shows how the conductance of an AGNR varies in 
general with the Fermi energy.

In view of the particle-hole symmetry of this problem, we show only the results 
for the conductance at positive energies. The value of the full transport gap is 
$E_g=2\Delta_{\varepsilon}$, where $\Delta_{\varepsilon}$ is the transport gap 
observed in plots such as the ones in \Fref{fig:Gjunction}, where only 
positive energies are shown. Throughout this work we shall refer to 
$\Delta_{\varepsilon}$ as the ``transport gap'', being clear that it corresponds 
to half of $E_g$.
The specific parameters for this system are 
$W=15.2$\,nm (ribbon width), $\varepsilon=0.03$ (strain magnitude), 
$\theta=\pi/2$ (strain direction), and the onset of strain is smoothed over 
different lengths, $L_s$. It is evident that, although the GNR is metallic in
the absence of any strain, a transport gap $\Delta_{\varepsilon}$ develops, and 
its magnitude ($\Delta_{0.03}=0.0435t_0$) is insensitive to $L_s$. This means 
that, as far as the existence and magnitude of the transport gap at low 
energies is concerned, smoothing of the unstrained-strained interface has no 
effect (in the absence of disorder). The effect is therefore robust with 
respect to the degree of sharpness of the strain barrier, which is an important 
result since a real strain interface will never be abrupt, and strain will 
always develop incrementally.
This is similar to what happens in a quantum point contact: when the width 
of the constriction is reduced gradually (adiabatically), the inter-subband 
mixing is reduced and the accuracy of the conductance quantization markedly 
improves \cite{deJong}. One consequence of this is that, as 
seen in \Fref{fig:Gjunction}(a), the smoother the junction, the better defined 
the quantization plateaux become. In this sense smoother junctions are not only 
a necessity imposed by the actual elastic behavior of the system and 
experimental conditions, but also a desirable situation as far as observation of 
conductance quantization is concerned.
\Fref{fig:Gjunction}(b) shows the typical behavior of the conductance of a 
smooth strain junction with increasing strain magnitude: the transport
gap increases with $\varepsilon$ ($\Delta_{0.06}=0.087t_0$), but the
conductance quantization remains unaffected, since plateaus are created at 
integer values of $G_0$, with strain changing only the energies of the 
conductance steps. 

It is crucial not to confuse the appearance of the transport gap with the 
finite sub-band spacing in a narrow nanoribbon. In other words, one could be 
tempted to think that, since the strained region is still a nanorribon of the 
same width but with the sub-bands slightly rearranged in energy, one would 
expect a small transport gap if the band arrangement in the strained region is 
such that it corresponds to an insulating nanoribbon. Even though this is 
true, it leads only to a small ``intrinsic'' gap scaling with $\sim 1/W$, which 
is rapidly overshadowed by the strain-induced gap, whose scale is set by 
$\varepsilon$ and is largely insensitive to the width $W$.

\begin{figure}[t]
  \centering
  \includegraphics[width=8.0cm]{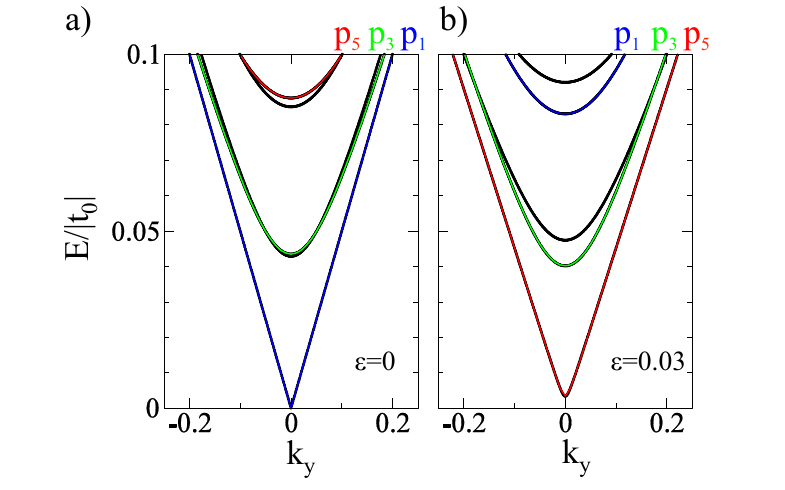}
\caption{
    (Color online) Band structure of an unstrained (a) and strained (b) AGNR 
    with $W=15.2$\,nm ($\mathcal{W}=125$, $\varepsilon=0.03$ and 
    $\theta=\pi/2$). 
    The first, third, and fifth conduction bands of the unstrained nanoribbon 
    are numbered and their new position in the strained nanoribbon is 
    highlighted.
  }
  \label{fig:bandas}
\end{figure}

In order to understand the behavior of the conductance reported in 
\Fref{fig:Gjunction} it is instructive to review the band structure of the 
system and consider the characteristics of the incoming, reflected, and 
transmitted modes. The band dispersion of an AGNR with $\theta=0$ or 
$\theta=\pi/2$ (i.e., $t_1=t_3$) is given by $E(q_x,k_y)=\pm |\phi(q_x,k_y)|$, 
where
\begin{equation}
  \phi= t_2e^{-ik_ya}
  +2t_{13}e^{\frac{ik_ya}{2}}\cos(\frac{\sqrt{3}a}{2}q_x)
  \label{eq:eqmod}
\end{equation}
and $t_{13}=t_1=t_3$. The transverse momentum is quantized as 
$q_x=\frac{2\pi}{\sqrt{3}a(\mathcal{W}+1)}\,p$, where $\mathcal{W}$ is the 
number of sites along a zigzag line in the transverse direction. 
The $\mathcal{W}$ possible values of $q_x$ are identified by the mode index 
$p=1,2,\hdots \mathcal{W}$. For a given energy and transverse momentum $q_x$ 
there are two possible values of longitudinal momentum $k_y$. All real values of 
$k_y$ for a given $q_x$ constitute a one-dimensional sub-band and, hence, the 
mode index $p$ also labels the sub-bands.

If we use $i$ to enumerate the conduction sub-bands in terms of increasing 
energy at $k_y=0$ we can then refer to their respective mode numbers as $p_i$. 
In \Fref{fig:bandas}(a) we have an example of this notation: the 
first band above $E=0$ is the one corresponding to the mode $p_1$, the second 
to the mode $p_2$, and so on. In addition, since our unstrained ribbons are 
always metallic, we can use the fact that $\mathcal{W}$ can be cast as 
$\mathcal{W}=3m+2$, with $m$ a positive integer, and obtain the mode 
number of the lowest sub-band: $p_1=2m+2$ \cite{Zheng}. 
The usefulness of this labeling will now be made apparent with a particular 
example. 

\subsection{Transport gap and conductance quantization}

\begin{figure}[t]
\centering
\includegraphics[width=8cm]{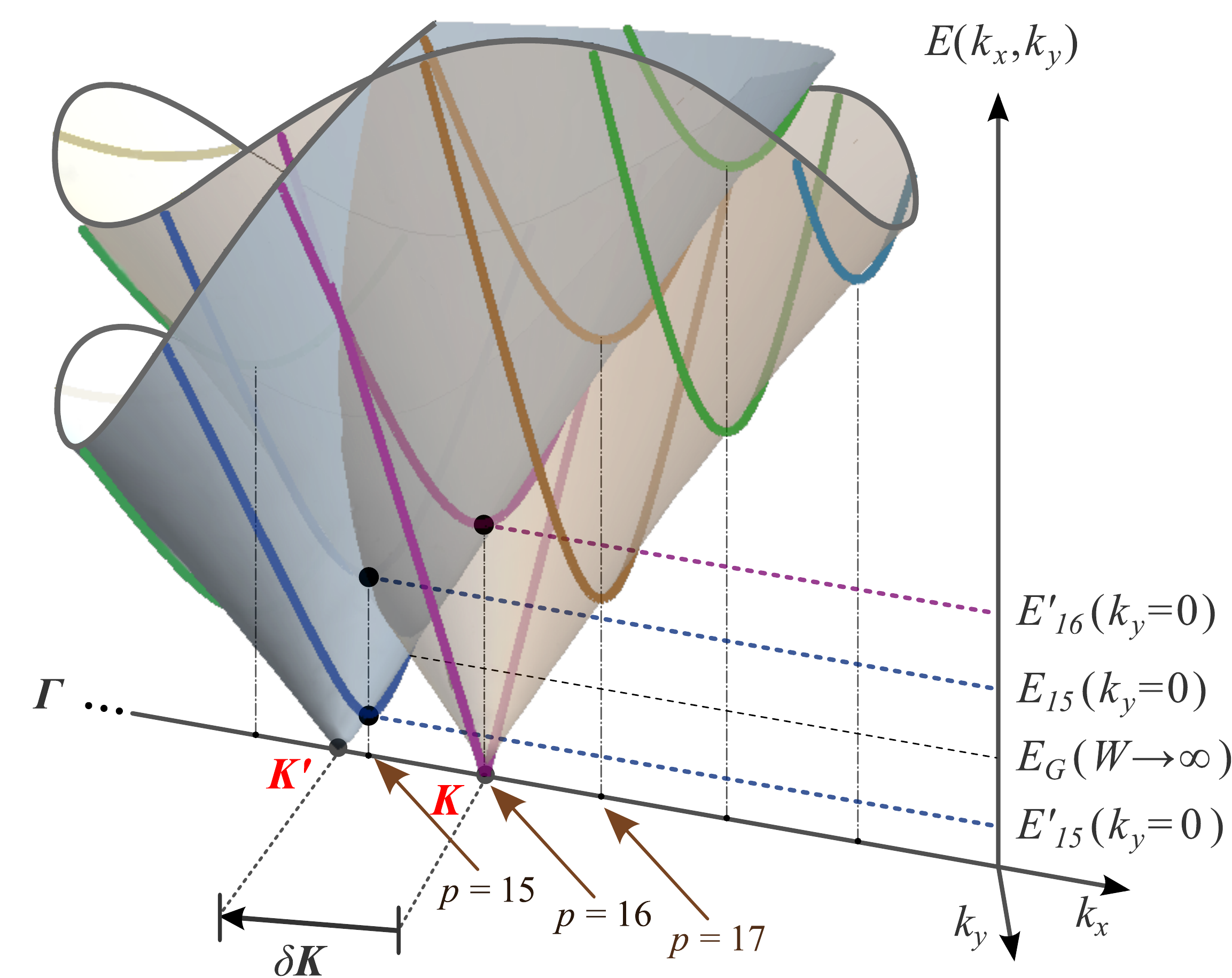}
\caption{
(Color online)
Illustration of the sub-bands of a GNR as ``slices'' of the 2D dispersion    of 
graphene at equidistant values of the transverse momentum (here     
$\mathcal{W}=23$). In 2D graphene the effect of uniaxial strain is to    
displace the Dirac point by $\delta\bm{K}$ in reciprocal space. Since the 
momentum quantization is determined solely by the width and chirality, the 
quantized momenta will be the same in the strained and unstrained    regions. 
Consequently, sub-bands corresponding to the same transverse momentum (or mode 
number) will appear at different energies in the strained region, when compared 
to the unstrained one. In this example the transport gap would be given by 
$E_{p=15}(k_y=0)$, and its thermodynamic limit when $W\to\infty$ is noted as 
$E_G$.
}
\label{fig:cones}
\end{figure}

Consider \Fref{fig:bandas} where the lowest sub-bands of a particular AGNR are 
plotted, with and without strain. Strain naturally modifies the electronic 
structure of the AGNR by (i) creating, in general, a small $W$-dependent 
intrinsic gap ($0.0033t_0$ in this particular case), and (ii) by reordering the
relative position and energy of the sub-bands corresponding to the same 
mode index. In \Fref{fig:bandas}(a) we highlight the first, 
third, and fifth bands of the unstrained AGNR (whose mode indices are $p_1=84$, 
$p_3=83$ and $p_5=82$), and show in \Fref{fig:bandas}(b) how the bands with the 
same mode numbers appear at different energies. It can be seen that the fifth 
band becomes the lowest energy band of the strained AGNR and the first band of
the unstrained AGNR becomes the fourth. \Fref{fig:cones} illustrates this band 
rearrangement from a perhaps more instructive perspective.
To understand why tracing the mode indices in the unstrained and strained 
regions is significant we should recall that, as stated above, the mode index 
defines the transverse momentum via 
$q_x=\frac{2\pi}{\sqrt{3}a(\mathcal{W}+1)}\,p$. Since strain varies only 
along the longitudinal direction, the transverse momentum (or mode index) should 
be preserved across the interface.
Additionally, since our geometry assumes an identical ribbon width in both 
strained and unstrained regions, the value of $q_x$ is identical on both.
This means that an incoming mode $p_k$ will only propagate across the interface 
if that mode is ``open'' in the strained region, which is to say, if the 
corresponding band lies above $E_F$.
This is precisely what happens, and what determines the transport gap. 
Inspecting the energy eigenvalues at the $\Gamma$ point of the strained AGNR it 
is found that the value of the transport gap [$\Delta_{0.03}=0.0435t_0$, cf. 
\Fref{fig:Gjunction}] does not tally with an eigenenergy of the strained AGNR, 
but, instead, it coincides with an eigenenergy of the unstrained AGNR 
corresponding to the third band ($p_3=83$). In other words, as $E_F$ is 
increased from zero the incoming electrons belong to mode $p_1$ up to 
$E_F=0.0435t_0$; this mode can only propagate in the strained region for 
$E_F\gtrsim0.08 t_0$; hence the conductance is zero. The lowest common band to 
the unstrained and strained region is the one associated with the mode $p_3$. 
Only when $E_F$ is increased past the minimum of this band can we have mode 
conservation, and this is what determines the transport gap.

\begin{figure}[t]
  \centering
  \includegraphics[width=8cm]{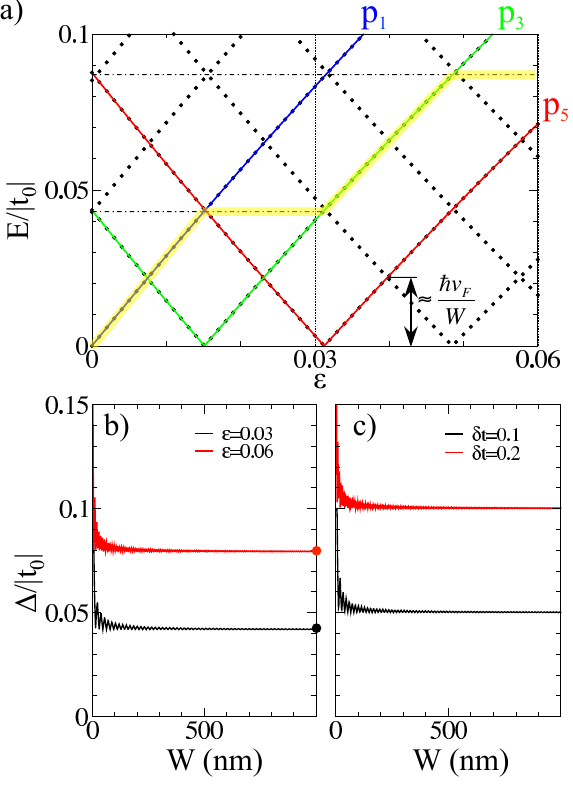}
\caption{
(Color online) (a) Evolution in energy of the modes at the
$\Gamma$ point as a function of $\varepsilon$ for the AGNR of $W = 15.2$\,nm 
and $\theta=\pi/2$. 
The evolution of the first (blue), third (green) and fifth (red) mode has 
been highlighted as well as the intrinsic gap $\approx \hbar v_F/W$. The yellow 
shaded region indicates the evolution of $\Delta_{\varepsilon}$.
(b) Transport gaps $\Delta_{0.03}$ and $\Delta_{0.06}$ as a function of ribbon 
width for $\varepsilon=0.03$ and $0.06$, respectively. The  black (red) dot in 
the right vertical axis marks the value of the transport gap $\Delta_{0.03} 
\approx 0.04t_0$ ($\Delta_{0.06} \approx 0.08t_0$) calculated using Dirac's 
equation. (c) Transport gap $\Delta_{\delta t}$ as a function of width for 
$\delta t=0.1$ and $\delta t=0.2$, where $t_2=t_0-\delta t$. The values of 
$\delta t$ were chosen to approximate the values of $t_2$ for $\varepsilon=0.03$ 
and $\varepsilon=0.06$.
}
  \label{fig:evolmod}
\end{figure}

To delve more into this point we can follow the evolution of the eigenenergies 
at $k_y=0$ as a function of strain. The results are plotted 
in \Fref{fig:evolmod}(a), where the evolution of the energies for modes $p_1$, 
$p_3$, and $p_5$ is highlighted. The variation of the eigenenergies is linear 
with strain, which is expected in connection with the illustration of 
\Fref{fig:cones}: first, the displacement of the ``enveloping Dirac cone'' is 
linear in strain \cite{Ando,VPereira3,netoRMP2009}, and, second, the energy is 
linear in the momentum close to the Dirac point.

As a result of the variation of the individual sub-bands with energy the 
intrinsic gap in the strained ribbon oscillates as strain increases, with an 
amplitude $\approx \hbar v_F/W$; this is simply a consequence of the 
displacement of the Dirac cone induced by strain, and the constancy of the 
quantized transverse momenta \cite{Yang, Nagapriya, Lu, Sun,Li}. The transport 
gap $\Delta_{\varepsilon}$, however, increases steadily with strain (except for 
a sawtooth modulation $\approx \hbar v_F/W$). This can be appreciated with the 
aid of \Fref{fig:evolmod}(a) that shows the minimum of each band as a function 
of strain. For example, following the vertical dashed line at 
$\varepsilon=0.03$ the first eigenenergy is $0.0033t_0$; this value corresponds 
to mode $p_5=82$. Since this mode is evanescent in the unstrained contact there 
is no transmission. The second eigenenergy is $0.0402t_0$; this value 
corresponds to mode $p_3=83$. For this energy this mode is evanescent in the 
unstrained
contact and consequently there is no transmission either. When the Fermi
energy reaches the first horizontal dashed line, that is when the mode
$p_3$ becomes propagating in the unstrained contact and starts to
transmit, the transmission gap $\Delta_{0.03}=0.0435t_0$ is reached
since the modes $p_3=83$ are opened in both unstrained and strained
contacts. Increasing the Fermi energy to the value of the eigenenergy
$0.0474t_0$ corresponding to mode $p_7=81$, there is no effect on the
conductance because for this energy the $p_7$ mode is evanescent in
the unstrained contact. Such a fine-grained analysis can be used to 
understand the position and width of the conductance plateaus in 
\Fref{fig:Gjunction}. Consider, for example, the very narrow $2G_0$ plateau for 
$\varepsilon=0.03$ in \Fref{fig:Gjunction}. It appears when the Fermi energy 
reaches the fourth eigenvalue ($0.0831t_0$) of the strained contact, and both 
modes (in the unstrained and strained regions) with $p_1=84$ are propagating 
and can transmit. Finally when the Fermi energy comes to the second horizontal
dashed line $0.0876t_0$ the mode $p_5=82$ of the unstrained contact,
finally is propagating and there is transmission in that mode. This
also explains the origin of the shorter plateaus observed in the
conductance of \Fref{fig:Gjunction}. Following the same procedure
for $\varepsilon=0.06$ the transmission gap $\Delta_{0.06}=0.0876t_0$
is extracted form \Fref{fig:evolmod}(a), it can be seen that the
transmission gap is created when the modes labeled with $p_5=82$ are
propagating in both contacts. 

Simply generalizing this reasoning and procedure we can trace the dependence 
of the transport gap for any value of strain  ($\varepsilon$)
and width ($W$). \Fref{fig:evolmod}(b) shows such results for $\Delta_{0.03}$ 
and $\Delta_{0.06}$ as a function of the width of the nanoribbon.
It can be seen that the transport gap has strong oscillations for narrow ($< 
100$\,nm) nanoribbons that rapidly die off with increasing $W$, on account 
of the reduction of spatial confinement in the transverse direction. For wide
junctions the transport gap eventually converges at the asymptotic values of 
$\Delta_{0.03}=0.042t_0$ and $\Delta_{0.06}=0.079t_0$, for the particular 
values of strain considered in the figure for illustration. These results
are in complete agreement with ones calculated using Dirac's equation
\cite{Pellegrino}. 

In addition, in order to make a direct 
comparison with the predictions for the transport gap predicted for 
the 2D graphene system under uniaxial strain, we calculated the transport gap 
$\Delta_{\delta t}$ of an unstrained-strained junction where only the hopping 
$t_2=t_0-\delta t$ is modified in the strained contact. The values of $\delta 
t$ were chosen to be equivalent to the values of $t_2$ obtained with 
$\varepsilon=0.03$ and $\varepsilon=0.06$. In this ideal situation it
is expected that $\Delta_{\delta t}=\delta t/2$ \cite{VPereira3} and, indeed,  
observing \Fref{fig:evolmod}(c) it can be seen that this is the 
asymptotic value for wide nanoribbons.

\subsection{Work function mismatch}

\begin{figure}[t]
  \centering
  \includegraphics[width=7.5cm]{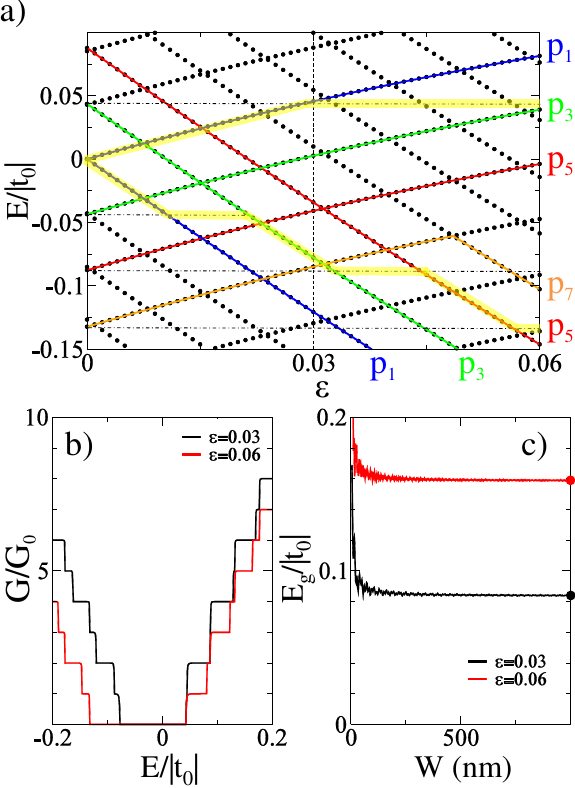}
\caption{
(Color online) (a) Evolution in energy of the modes at the
$\Gamma$ point as a function of $\varepsilon$ for the AGNR of $W = 15.2$\,nm
and $\theta=\pi/2$, including the work function mismatch as a scalar potential.
The evolution of the positive and negative branches of the first (blue), third
(green), and fifth (red) modes has been highlighted, as well as the negative 
branch of the seventh mode (orange).
For positive energies the yellow shaded region indicates the evolution of
$\Delta^{+}_{\varepsilon}$, while for negative energies it indicates the 
evolution of $\Delta^{-}_{\varepsilon}$.
(b) Conductance of an unstrained-strained junction with $W=15.2$\,nm and 
$\theta=\pi/2$, including the effect of the work function mismatch.
(c) Transport gap $E_g=\Delta^{+}_{\varepsilon}+|\Delta^{-}_{\varepsilon}|$
as a function of ribbon width for $\varepsilon=0.03$ and $\varepsilon=0.06$,
respectively. The black and red dots on the right vertical axis mark the value 
of the transport gap calculated without the work function mismatch:
$2\Delta_{0.03} = 0.084t_0$ and $E_g=2\Delta_{0.06} = 0.158t_0$ (see 
\Fref{fig:evolmod}b).
}
  \label{fig:evolMWF}
\end{figure}

To  efficiently inject carriers from one material to another a good band
alignment is required. The work function difference determines how the bands of 
different materials align when they are put in contact. It has been shown that 
the work function of graphene can be engineered through chemical doping 
\cite{WF_eng} or strain \cite{WF_strain,AGNR_WF}. In particular, strain is 
known to increase or decrease the work function depending on whether 
the lattice is, respectively, strained or compressed. Consequently, different 
work functions in the strained and unstrained regions lead to band misalignment 
and this appears to require a reformulation of the model and calculations 
presented in the previous section. The band mismatch created by homogeneous 
strain can be regarded as an effective scalar potential \cite{WF_strain} 
relating the Fermi energies in the unstrained and strained region as 
$E^s_F=E^{us}_F-(\phi_s -\phi_{us})$, where $\phi_{us}$ ($\phi_{s}$) is the 
work function in the unstrained (strained) region. For AGNR it has been 
found that the work function increases linearly with uniaxial tensile strain up 
to 12\,\%, regardless the width of the nanoribbon \cite{AGNR_WF}, and with a 
magnitude that can depend on the details of edge passivation. Since we are 
interested in the general consequences of a work function mismatch to the 
transport gap and conductance quantization, without compromising this 
generality we extracted $\phi_{us}=4.2$\,eV and 
$\phi_{s}(\varepsilon=0.04)=4.35$\,eV. These values correspond to the 
mismatch predicted in Ref.~\onlinecite{AGNR_WF} for hydrogen-passivated edges.
With these parameters, the Fermi energy in the strained region, $E_F^s$, can be 
written in terms of its unstrained counterpart, $E_F^{us}$, and strain 
magnitude, $\varepsilon$, as
\begin{equation}
E^s_F=E^{us}_F -3.75\varepsilon
\qquad\text{(eV)}
.
\label{eq:EF_WF}
\end{equation}
According to this, the effect of the strain-induced work function mismatch can 
be modeled by adding a strain-dependent on-site energy of $-3.75\varepsilon$
to the sites in the strained region. The addition of this effective scalar
potential does not affect the methods used in the previous section since 
transverse momentum conservation is still valid (the scalar potential is a 
function of strain, and this varies only along the longitudinal direction). We 
therefore used this approach to model and investigate the effects of work 
function mismatch. Since the strain variation is smoothed according to 
\Eqref{eq:smooth}, the effective local potential will vary smoothly as well.

The effect of incorporating explicitly this work function mismatch in our 
conductance calculations is a downwards displacement of the eigenenergies
in the strained region which, consequently, modifies the quantization 
signatures in the conductance trace, as well as the transmission gap. Since the 
conductance trace $G(E)$ is no longer particle-hole symmetric 
(\Fref{fig:evolMWF}b), the transport gap is now determined by 
$E_g=\Delta^+_{\varepsilon} + |\Delta^-_{\varepsilon}|$, which requires the 
explicit calculation of the transmission threshold at positive and negative 
energies, $\Delta^{\pm}_{\varepsilon}$. For example, resorting to 
\Fref{fig:evolMWF}a that shows the evolution of the different modes with strain, 
we see that $\Delta^+_{0.03}=0.0435t_0$ and $\Delta^-_{0.03}=-0.0777t_0$, 
leading to a transport gap $E_g=0.1212t_0$ that is larger than the value 
$2\Delta_{0.03}=0.087t_0$ obtained without work function mismatch. A larger 
transmission gap is also observed for $\varepsilon=0.06$ 
($E_g=(0.0435+0.132)t_0=0.1755t_0$). In this case, the asymmetry induced by the 
local potential has the consequence that the modes responsible for the gap have 
changed, and $\Delta^+_{0.06}=\Delta^+_{0.03}$ is determined by mode $p_3=83$ 
from the positive energy branch, while $\Delta^-_{0.06}=-0.132t_0$ is defined by 
mode $p_7=81$ from the negative branch. 
These considerations are confirmed by a direct calculation of the conductance, 
which is shown in \Fref{fig:evolMWF}b for these two particular values of strain 
(compare with \Fref{fig:Gjunction}b). The width dependence of the gap is shown 
in \Fref{fig:evolMWF}c. The asymptotic values for the full gap of 
$E_g=0.084t_0$ and $E_g=0.16t_0$ at, respectively, $\varepsilon=0.03$ and 
$0.06$, are seen to be equal to the full gaps obtained without work function 
mismatch (see \Fref{fig:Gjunction}c). This is highlighted by the dots located on 
the right vertical axis, which mark the positions $E_g=2\Delta_{0.03}=0.084t_0$ 
and $2\Delta_{0.06}=0.158t_0$, where $\Delta_{\varepsilon}$ are the ones 
calculated in \Fref{fig:Gjunction}c. 

The main message from here is that the work function mismatch displaces the 
center of the gap from $E=0$ to 
$E=(\Delta^+_{\varepsilon}+\Delta^-_{\varepsilon})/2 \approx 
-3.75\varepsilon/2$, but its magnitude remains unchanged relative to the case 
where work function mismatch is disregarded. This is a direct consequence 
of the linear spectrum of graphene which causes the strain-induced transport 
gap in the 2D limit to be independent of a uniform, but different, potential 
energy in the two regions. Moreover, the analysis of the mode 
evolution with strain from in \Fref{fig:evolMWF}a shows that the finite 
transport gap is still a consequence of the conservation of mode index 
(transverse momentum), and that the gap is determined by the lowest energy 
modes that are simultaneously ``open'' in the strained and unstrained 
regions. Finally, since we are ultimately interested in the behavior of the 
transport gap as a function of strain, these facts allow us to concentrate 
only on strain junctions where the explicit work function mismatch is ignored 
without losing generality as far as the magnitude and strain-dependence of the 
gap is concerned. In the remainder of this paper we therefore will not 
consider explicitly the work function mismatch.

\subsection{Mode mixing}

\begin{figure}[t]
  \centering
  \includegraphics[width=7.5cm]{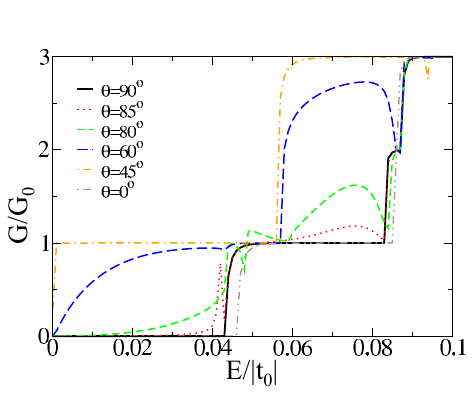}
\caption{
(Color online) Conductance of an unstrained-strained junction
($W=15.2$\,nm) with $\varepsilon=0.03$ for different directions of 
uniaxial tension $\theta$.
  }
  \label{fig:evolang}
\end{figure}

When the strain is not in the transverse or longitudinal directions
($\theta \neq 0,\, \pi/2$), \Eqref{eq:eqmod} is no longer valid to calculate the 
dispersion relation, consequently the values of the quantized transverse 
momentum in the strained contact will differ form the values of the quantized  
transverse momentum in the unstrained contact. This will lead to mode mixing,
even for a smooth junction or a junction without any kind of
disorder. An electron incident upon the junction in a given mode will be
mixed with a number of modes of the strained contact with similar
transverse momentum
and symmetry of the wave function \cite{PRL_wfsym,wang_wfsym}
, as a consequence the transport gap will be
reduced or disappears since its existence was due to the matching of
the lowest energy propagating modes. Mode mixing will be small while
the angle does not deviate too much from ideal situations ($\theta =
0$ and $\theta =\pi/2$), but this effect will reduce the value of the transport
gap, as can be seen in \Fref{fig:evolang} where the conductance of
a strained junction of $W=15.2$\,nm with $\varepsilon =0.03$ is plotted
for different angles of the applied strain. For $\theta=85^{\circ}$ we
obtained a transport gap $\Delta_{0.03}=0.04t_0$, and for $\theta=80^{\circ}$
we have $\Delta_{0.03}=0.027t_0$. When the band structure is
calculated for the strained AGNR, using the latter values, it can be
corroborated that, in fact, there exists a transport gap since there are lower
energy propagating modes that are not transmitting. For $\theta =60^{\circ}$ one 
sees an apparent transport gap $\Delta_{0.03}=0.002t_0$, but it is only apparent 
because when the band structure is calculated this value corresponds to the 
lowest energy propagating mode and, hence we are observing the intrinsic gap, 
not a transport gap. When $\theta=45^{\circ}$ there is no transport or intrinsic 
gap. Observing \Fref{fig:evolang} it is clear that any amount of mode mixing 
destroys the conductance quantization: some lineshapes of the conductance at 
higher energies can be traced back in the band structure of the strained 
contact. However for low energies the effect is completely due to the degree of 
mode mixing induced by angle of the applied strain. 

\begin{figure}[t]
  \centering
  \includegraphics[width=8.5cm]{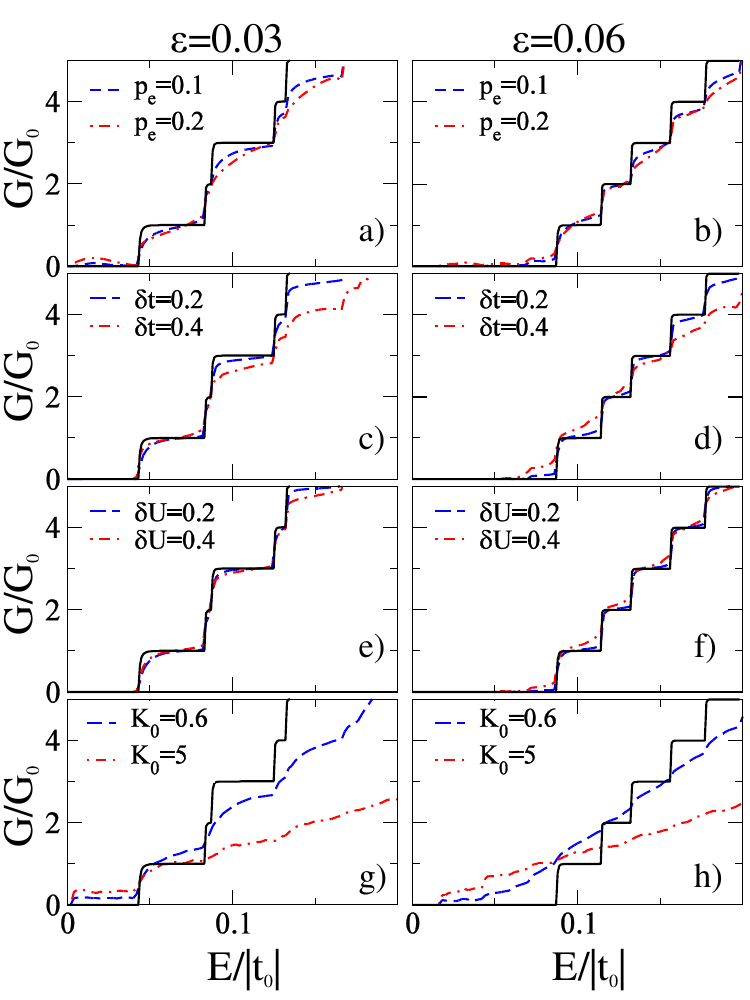}
  \caption{
    (Color online) Average conductance of a disordered unstrained-strained 
junction ($W=15.2$\,nm, $\theta=\pi/2$). The length of the disordered
region is $L=W=15.2$\,nm with $\varepsilon=0.03$ (left column) and
$\varepsilon=0.06$ (right column). The disorder models shown are:
(a,b) edge disorder with $p_e=0.1$ and $p_e=0.2$,
(c,d) hopping disorder with $\delta t=0.2$ and $\delta t=0.4$,
(e,f) on-site short range disorder with $\delta U=0.2$ and $\delta U=0.4$,
(g,h) on-site long range disorder with $K_0=0.6$, $n_{imp}=0.02$, $\xi=3a_0$, 
and $K_0=5$, $n_{imp}=0.04$, $\xi=3a_0$,
  }
  \label{fig:jundis}
\end{figure}

Another way to induce mode mixing is by adding disorder to the problem. To 
be specific, we consider the effect of adding disorder in a region of length 
$L=W$ between the perfect unstrained and strained contacts. 
Four disorder models were analyzed: (i) edge disorder, (ii) hopping 
disorder, (iii) short range bulk disorder, and (iv) long range bulk disorder. 
Edge disorder was implemented by removing sites from the outermost row of 
carbon atoms with a probability $p_e$. In the hopping disorder model every 
hopping at the disordered region was modified according to 
$\tilde{t}_{ij}=t_{ij}+\Delta t$, where
$\Delta t$ is uniformly distributed in the interval $[-\delta
t/2,\delta t/2]$. The short-range bulk disorder was modeled via a
random on-site energy $u_i=\delta u$, where $\delta u$ is a random
number uniformly distributed over $[-\delta U/2,\delta U/2]$. Finally, long
range bulk disorder was generated by randomly distributing $N_{imp}$ 
impurities, each modeled by a Gaussian function of width $\xi$, and with an 
amplitude $U_n$ selected from the uniform distribution $[-\delta 
U_g,\delta U_g]$. These impurities cause a modification of the on-site energy 
given by
\begin{equation}
  u_i=\sum^{N_{imp}}_{n=1}{U_n\, e^{-(r-R_n)^2/2\xi^2}} 
\end{equation}
and its strength is quantified with the dimensionless  parameter 
$K_0$, which in the dilute limit can be expressed as $K_0 \approx 
40.5n_{imp}(\delta U_g/t_0)^2(\xi/a_0)^4$, where $n_{imp}$ is the impurity 
density and $a_0$ is the lattice constant. Our strategy was to introduce a 
small amount of disorder to see how, for a given strain 
magnitude $\varepsilon$, the transport gap $\Delta_{\varepsilon}$ and the 
conductance were modified. The result is shown in \Fref{fig:jundis}, where we 
show the conductance averaged over 100 disorder realizations for each 
model and for two values of strain $\varepsilon=0.03$ (left column) 
and $\varepsilon=0.06$ (right column). Focusing on the transport gap, it can be 
seen that it is more strongly affected by edge and long-range disorder, 
its suppression being largely insensitive to the disorder strength in both 
situations. On the one hand, edge defects will induce strong back-scattering, 
particularly in low energy modes of the  unstrained contact \cite{Mucciolo, 
Evaldsson,TCLi,Zarbo} which are just the modes reflected at the clean 
unstrained-strained junction. This back-scattering leads to mixing with the 
propagating modes below $\Delta_{\varepsilon}$ in the strained contact and, 
consequently, the transmission gap disappears. This effect is reduced in 
wider junctions (not shown), as well as in unstrained AGNR \cite{Mucciolo, 
Evaldsson}. On-site long range disorder, on the other hand,  will lead
electron-hole puddles that strongly impact the low energy states 
\cite{Martin}. This again will introduce mode mixing between the low energy 
modes of the unstrained contact with the low energy propagating modes below 
$\Delta_{\varepsilon}$ in the strained contact. Turning our attention now to the 
other two types of disorder --- hopping in \Fref{fig:jundis}(c,d) and short 
range in \Fref{fig:jundis}(e,f) --- one sees that mode mixing is moderate since 
the transport gap is only reduced, and the conductance quantization is still 
preserved to a very good degree \cite{Klos,Aurelien}. Comparing the 
left and right panels of \Fref{fig:jundis} one can see that the same amount of 
disorder washes out more efficiently the conductance features  of the junction 
under the highest strain. This can be explained as arising from the fact 
that the larger the strain the more high energy modes of the unstrained
contact become low energy propagating modes of the strained contact, as shown in 
\Fref{fig:evolmod}(a). Therefore, in the energy range under consideration here, 
these modes can only by conducting because of the mode mixing.

\section{Conductance of a disordered strained barrier}

\begin{figure}[t]
  \centering
    \includegraphics[width=8.5cm]{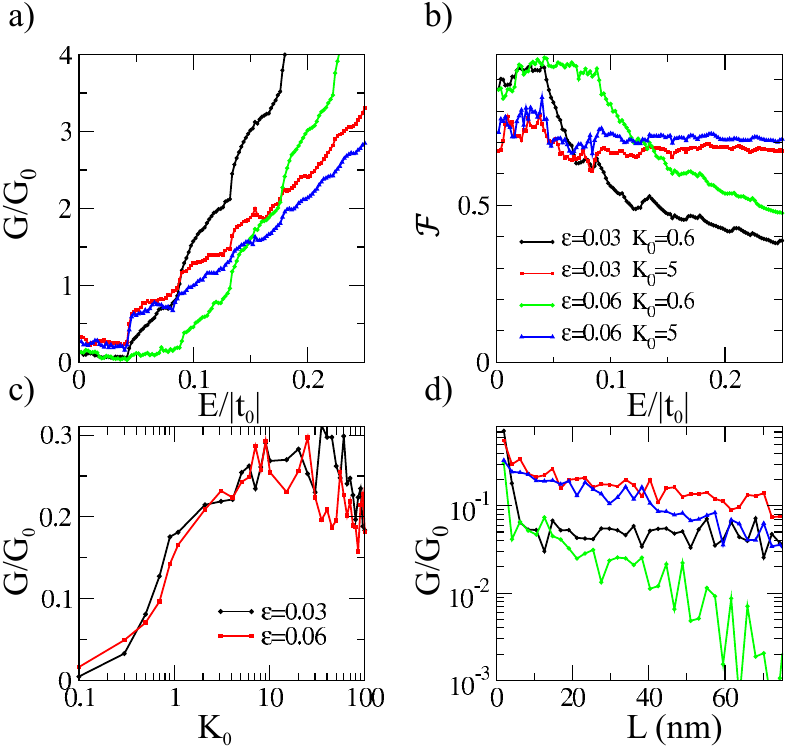}
  \caption{
    (Color online) Average conductance (a) and average
Fano factor (b) of a disordered strained barrier ($W=L=15.2$\,nm, 
$\theta=\pi/2$) for $\varepsilon=0.03$ and $\varepsilon=0.06$ in the
presence of on-site long-range disorder characterized by \{$K_0=0.6$, 
$n_{imp}=0.02$, $\xi=3a_0$\} and \{$K_0=5$, $n_{imp}=0.04$, $\xi=3a_0$\}. 
(c) Average conductance for $E=0.01t_0$ as a function of disorder strength
of the same barrier studied in (a,b). The potential was smoothed over $\xi=3a_0$ 
for different values of the impurity density $n_{imp}$.
(d) Average conductance for $E=0.01t_0$ as a function of the length of the 
disordered region ($W=15.2$\,nm, $\xi=3a_0$, and different values of the 
impurity density $n_{imp}$ have been used, keeping $K_0=0.6$ and $K_0=5$).
  }
  \label{fig:Gbar}
\end{figure}

We now focus on the conductance of a disordered strain barrier. As defined 
earlier, this system consists of a strained region ($\theta=\pi/2$) of length 
$L=W$ between two perfect unstrained semi-infinite contacts (as 
illustrated in \Fref{fig:fig1}(b). In the absence of disorder and when the 
strain barrier is smooth the resulting conductance is exactly the same
to that observed in the unstrained-strained junction. 

This is expected because electron transmission can only occur when a given 
mode is opened in the three regions. Left and right contacts are exactly the 
same, and so the same mode will be accessible at the same energy in both 
contacts. In this case the reasoning of \Fref{fig:evolmod}(a) can be repeated 
resulting in the same quantum conductance. The situation is analogous to the 
quantum point contact example: the quantization of the conductance is given by 
the narrower constriction.

Since we established above that on-site long-range disorder is the most 
efficient mode mixer, this is the only disorder model that we report on now. 
The resulting conductance averaged over 100 disorder realizations is presented 
in \Fref{fig:Gbar}(a). There is no transport gap in the average conductance 
because of the relatively strong mode mixing in the strained disordered 
barrier (and there is no intrinsic gap in the contacts either). For low 
levels of disorder, $K_0=0.6$, it is evident that mode mixing is lower since the 
lineshape of  average conductance bears some resemblance to that of the 
unstrained-strained junction. It is even possible to observe the appearance 
of the small plateaus that were discussed above. It can also be concluded
that larger strain transmits fewer modes (the conductance for a given 
energy is lower), and that the value of the conductance is small ($G
\approx 0.1G_0$) in the energy range that corresponds to
$\Delta_{0.03}$ and $\Delta_{0.06}$ of the unstrained-strained
junction. When the disorder strength increases ($K_0=5$)  there is no
appreciable difference between the average conductance with
$\varepsilon=0.03$ and $\varepsilon=0.06$, and, in fact, the overall features 
of the conductance traces are no different from the conductance of a disordered 
unstrained barrier \cite{Mucciolo,Ihnatsenka} (but still, there is an increase 
of the average conductance, and the clear definition of the first two 
plateaus with values $G \approx 0.3G_0$ and $G \approx 0.7G_0$).  

Shot noise is quantified by the dimensionless Fano factor
$\mathcal{F}$, in particular it is a quantity that reveals information
about the transport dynamics in the device.
$\mathcal{F}=\sum_pT_p(1-T_p)/\sum_pT_p$, where $T_p$ is the
transmission probability of mode $p$ \cite{YBlanter,deJong}. In the
eigenmode representation the device can be seen as a parallel circuit
of $\mathcal{W}$ independent transmission modes, to access this
representation in graphene -- without using the analytic expression
for metallic contacts \cite{Ferry,Schomerus,Lewenkopf,Dragomirova} --
it is necessary a numerical method \cite{Ando,Khomyakov,Hansen}. 
\Fref{fig:Gbar}(b) shows the average Fano factor, using the same values
of the average conductance of \Fref{fig:Gbar}(a). For low disorder
($K_0=0.6$) and energies in the range of $\Delta_{\varepsilon}$ the
Fano factor, $\mathcal{F} \approx 0.95$,  indicates a tunnel barrier
behavior this is $T_p \ll 1$, as can be expected since there is no
mode matching for that energy range. For the same energy range it is
seen that the Fano factor oscillates, a clear indication that mode
mixing is the main mechanism for the conductance enhancement. For
higher energies than $\Delta_{\varepsilon}$, the Fano factor begins to
decay, that because of the mode matching between propagating modes in
the contacts and in the strained barrier. For $K_0=5$ there is no
qualitative differences between the Fano factor for different values
of strain, disorder enhances the mode mixing and the transmission is
increased (Fano factor is smaller), especially in the energy range of
the first two plateaus of the average conductance in \Fref{fig:Gbar}(a), as can 
be seen in the higher oscillations of Fano
factor in that energy range. For higher energies the oscillations are
damped and the Fano factor saturates. We followed the evolution of
Fano factor for higher values of disorder up to $K_0=10$ (not showed)
finding no significant changes in the lineshape and values of the
average Fano factor.

Looking more closely the effect of disorder on the conductance and its
relation with strain, especially in the energy range of
$\Delta_{\varepsilon}$, we fixed the Fermi energy at $E=0.01t_0$ and
increase the impurity density ($n_{imp}$) in the strained barrier.
After averaging over 100 realizations the average conductance is
plotted in \Fref{fig:Gbar}(c), it can be seen that the effect of
strain is completely washed out by disorder, there is no appreciable
differences between the curves with $\varepsilon=0.03$ and
$\varepsilon=0.06$. With the increase of $K_0$ the conductance is
enhanced from $G \approx 0.01G_0$ for low disorder to $G \approx
0.3G_0$ in the intermediate disorder regime. We switch now to examine
the effect of length of the strained barrier in the conductance, again
we set the Fermi energy  to $E=0.01t_0$ and averaging over 500
disorder realization, the resulting conductance is plotted in 
\Fref{fig:Gbar}(d). It is observed that the average conductance decreases
exponentially with the length, for all disorder strength and applied
strain. For $K_0=5$ it could be said that the localization length is
roughly the same and that there are no appreciable differences in the
lineshape and values of the average conductance for different values
of strain. For $K_0=0.6$ the localization length is shorter for the
higher strain value, this effect will be appreciated in lower
conductance values for larger barriers, however there is no the
formation of a transport gap.

\section{Summary}

The electronic transport across strained junctions and barriers in a graphene 
nanorribon has been studied in the framework of Landauer and B\"uttiker, 
implemented using non-equilibrium Green's functions. A clear strain-dependent 
transport gap appears for strain applied along the zigzag and armchair 
directions of non-disordered strain junctions and barriers. The transport gap is 
a result of the perfect matching between the propagating modes in both regions. 
A different angle of applied strain or disorder induces mode mixing, which tends 
to degrade the gap as well as the conductance quantization, both signatures of 
the electronic transport across a strained region. For Fermi energies in the 
energy range of the transport gap, the presence of unmatched propagating 
modes in the strained region means that they become active in the presence 
of disorder, leading to a conductance plateau that is sustained for a broad 
range of disorder strength. The conductance in this plateau decays exponentially 
indicating that strain induces localization in the low energy single channel 
regime \cite{Yamamoto2}. 

We have shown results for disordered square junctions and barriers with 
$W=15.2$\,nm. However, our results can be easily extrapolated to junctions and 
barriers of different aspect ratios. For larger disordered regions the 
number of scattering centers grows, the conductance decays and the quantization 
plateaus are destroyed for energies higher than the transport gap of the clean 
junction. For energies below the transport gap the conductance is enhanced 
leading to a broad plateau whose conductance value can be smaller that $G_0$. 
Although not shown explicitly here, this was observed by direct calculations in 
junctions and barriers with $W=10$\,nm , $W=30$\,nm, and $W=50$nm, as well 
as different aspect ratios.

Our results complement the existing calculations based on the 2D electronic 
dispersion that predict a strain-dependent transport gap at low densities, and 
lend support to the idea that this gap is robust against strain inhomogeneity, 
barrier smoothness and some degree of electronic disorder.

%
\acknowledgments
This work was supported by the NRF-CRP award ``Novel 2D materials with
tailored properties: beyond graphene'' (R-144-000-295-281).

\bibliographystyle{apsrev}
\bibliography{conductance_strain}

\end{document}